\documentclass[cits]{PoS}
\usepackage{axodraw}
\usepackage{amsmath, amssymb}

\newcommand{\msbar}{{\rm{\overline{MS}}}}

\title{Phenomenology from the Lattice}

\ShortTitle{Lattice Phenomenology}

\author{\speaker{Christopher Sachrajda}%\thanks{A footnote may follow.}
\\
        School of Physics and Astronomy\\
        University of Southampton\\
        Southampton SO17 1BJ, UK\\
        E-mail: \email{cts@soton.ac.uk}}

\abstract{In recent years the precision of lattice calculations has improved hugely, and the results are making a very significant impact in particle physics phenomenology. Indeed there is no alternative general method which can be used in the evaluation of nonperturbative strong interaction effects for a wide variety of physical processes. In this talk I discuss a selection of topics in flavour physics, including \textit{mature} quantities for which lattice calculations have been performed for a long time (e.g. the determination of the $V_{us}$ CKM matrix element and $B_K$), quantities which we are now learning to study (e.g. $K\to\pi\pi$ decays amplitudes and the spectrum and mixing of $\eta-\eta^\prime$ mesons) and important phenomenological quantities for which a large amount of experimental data is available but which we do not yet understand how to approach in lattice simulations (e.g. nonleptonic $B$-decays). The improvement in precision and the extension of the range of processes which can be studied using lattice QCD has to be continued vigorously if precision flavour physics is to play a complementary role to large $P_\perp$ discovery experiments at the LHC in unravelling the next level of fundamental physics.}

\FullConference{The XXVIII International Symposium on Lattice Field Theory, Lattice2010\\
		June 14-19, 2010\\
		Villasimius, Italy}
\begin{document}

\section{Introductory Comments}\label{sec:intro}
Recent progress in lattice phenomenology has been truly impressive, due to improved algorithms and theoretical developments as well as
more powerful computing resources. It is clearly impossible to do justice to the title of this talk by discussing all the exciting new developments and results and I will not try to do this. At this conference there have been a number of excellent review talks on specific topics including those in refs.~\cite{Alexandrou:2010cm,Heitger,Herdoiza:2011gp,Hoelbling:2011kk,Laiho} as well as many parallel sessions in which the new ideas and results have been presented explicitly. I will therefore not attempt to make a systematic compilation of the latest lattice results. I will also not discuss the different formulations of lattice QCD or effective theories of heavy quarks. Instead I will discuss and give examples of
\begin{enumerate}
\vspace{-0.05in}\item \textit{Mature} quantities, i.e. ones which have been studied successfully and with improving precision for many years. These include the determination of the CKM matrix element $V_{us}$ and the $B_K$ parameter of neutral kaon mixing, which I will discuss. The natural question which arises for such quantities is \textit{what next?} Does the phenomenology require us to strive towards even higher precision or should we set our priorities towards the evaluation of other quantities?
\vspace{-0.05in}\item Continuing attempts to extend the range of physical quantities for which lattice simulations can contribute to the quantitative control of the nonperturbative QCD effects. In this context I will discuss attempts to evaluate $K\to\pi\pi$ decay amplitudes and a recent study of the $\eta-\eta^\prime$ system. I will also briefly mention relatively early attempts to evaluate long-distance effects, encoded in the matrix elements of time-ordered (non-local) products of operators.
\vspace{-0.05in}\item Quantities for which we don't yet know how to formulate the lattice calculation of the corresponding non-perturbative QCD effects. For illustration I will discuss two-body nonleptonic $B$-decays, a hugely important set of processes for which there is a large amount of experimental data and yet lattice simulations are not playing any significant r\^ole in the phenomenology.
\end{enumerate}
The examples I use in this talk are taken from Flavour Physics, but of course we also need to remember the contributions which lattice QCD is making to the development of our understanding of hadronic structure and to the determination of quark masses and the coupling constant $\alpha_s$. The \textit{mission} of precision flavour physics is to play a complementary r\^ole to large $p_\perp$ experiments in discovering and unraveling the next layer of fundamental physics beyond the standard model. If, as is expected, or at least hoped, the LHC experiments discover new elementary particles then precision flavour physics will be necessary to determine the underlying theoretical framework and lattice simulations will be central to this endeavour, with the specific task of quantifying nonperturbative hadronic effects. The discovery potential of precision flavour physics also should not be underestimated however, with a real possibility that current \textit{tensions} within the standard CKM analysis will be confirmed and become inconsistencies which will have to be explained by \textit{new physics}. Even restricting the discussion to flavour physics, I have to be selective and have naturally chosen to discuss the status of some topics which have been of direct interest to me in recent years, and which I believe are indicative of the exciting progress and prospects for the subject more widely.

Before proceeding to the discussion of specific processes, I start with some comments about the use of continuum perturbation theory which is necessary to relate renormalization schemes which can be simulated in lattice calculations and ones based on dimensional regularization (such as $\overline{\textrm{MS}}$) which cannot, but which are generally used in the perturbative evaluation of Wilson coefficient functions. The main point of this discussion is to underline that the precision of current lattice computations is such that, in order to get the maximum scientific benefit from the results we need to work with the higher-order perturbation theory community to obtain Wilson coefficient functions in schemes we can simulate. After this digression I proceed to a discussion of the status of the evaluation of a variety of physical quantities, starting with standard ones such as $V_{us}$ and $B_K$, through quantities which we are beginning to evaluate reliably such as $K\to\pi\pi$ decay amplitudes and $\eta-\eta^\prime$ masses and mixing and finally to some quantities which we currently do not know how to evaluate at all (non-leptonic $B$-decay amplitudes).

\subsection{Continuum Perturbation Theory}
Lattice simulations are used to compute the long-distance non-perturbative effects in quantities such as the matrix elements of local composite operators or the QCD parameters $\alpha_{~s}$ (the strong coupling constant) and the quark masses. These quantities require renormalization and so have to be combined with perturbative calculations before they can be used in predicting physical observables. This is sketched in the following oversimplified picture:
\begin{center}
\begin{tabular}{ccccc}\\
Physics&=&$C$&$\times$&$\langle\, f\,|\,O\,|\,i\rangle$\\
&&$\uparrow$&&$\uparrow$\\
&&Perturbative&&Lattice\\
&&QCD&&QCD
\end{tabular}\end{center}
where $C$ (perhaps a Wilson coefficient function) contains the short-distance physics and is calculated in perturbation theory and the long-distance effects are contained in the matrix element(s) of one or more local operators. The renormalization scheme and scale dependence cancels between the two factors $C$ and $\langle\, f\,|\,O\,|\,i\rangle$ provided that they are calculated in the same renormalization scheme. Because of its practical advantages $C$ is usually calculated in a scheme based on dimensional regularization, such as the $\msbar$ scheme. Unfortunately however, we are not able to perform lattice simulations in a non-integer number of dimensions and hence
$\msbar$, while being the scheme of choice in continuum perturbative calculations, is not directly useful in combining the perturbative and lattice results. In particular, it is not possible to calculate the matrix element in the $\msbar$ scheme entirely non-perturbatively.

One possibility for matching $C$ and the matrix element, as long as the ultraviolet cut-off ($a^{-1}$,~where $a$ is the lattice spacing) and the renormalization scale ($\mu$) are both sufficiently large, is to use perturbation theory to relate $\langle\, f\,|\,O^{\rm latt}(a)\,|\,i\rangle$, computed in the bare lattice theory and with bare operators $O^{\rm latt}(a)$, to the corresponding renormalized operators $O^{\overline{\rm MS}}(\mu)$. In addition to the technical difficulty of performing perturbative calculations in the lattice theory beyond one-loop order, it is found however, that the series frequently converge poorly. It is therefore preferable to perform {\em Non-Perturbative Renormalization} (NPR), by choosing a renormalization condition which can be imposed directly in lattice computations~\cite{Martinelli:1994ty} and examples of such schemes include the momentum schemes RI-Mom~~\cite{Martinelli:1994ty} and its recent generalization RI-SMom~\cite{Aoki:2007xm,Sturm:2009kb} and schemes based on the use of the Schr\"odinger Functional~\cite{Luscher:1992an}. Having obtained the matrix element in such an intermediate scheme, we need to combine it with $C$ calculated in the same scheme or equivalently to translate the matrix element from the intermediate scheme to $\msbar$. The precision of lattice calculations is now such that this translation from the intermediate scheme to $\msbar$ needs to be performed beyond one-loop order and hence is left most effectively to the professional N$^{\rm n}$LO perturbation theory specialists. It is therefore pleasing to see such calculations being performed; this enhances considerably the phenomenological reach of the lattice calculations. In addition to John Gracey, who has been performing such calculations for some time (see for example~\cite{Gracey:2003yr}), there have been recent calculations relating the quark masses in the RI-SMom and $\msbar$ schemes at two-loop order~\cite{Gorbahn:2010bf,Almeida:2010ns}. Note also the joint work by the HPQCD lattice collaboration and the Karlsruhe perturbation theorists on the evaluation of the charm-quark mass by matching moments of correlators on the lattice and in the continuum~\cite{Allison:2008xk}.

The principal lesson of this section is that close collaboration with the QCD perturbation theory community is increasingly necessary to optimise the significance of our lattice results.

\section{$V_{us}$ from Lattice simulations}
Lattice calculations of the decay constants $f_\pi$ and $f_K$ are by now standard and from the ratio $f_K/f_\pi$, combined with the experimental leptonic widths, we obtain the ratio $V_{us}/V_{ud}$. The current status obtained from simulations with $N_f=2+1$ flavours of sea quarks is~\cite{Colangelo:2010et}
\begin{equation}\label{eq:fkfpistatus}
\frac{f_K}{f_\pi}=1.193 \pm 0.006\,.
\end{equation}
In the last 6 years or so, following the suggestion of Becirevic et al.~\cite{Becirevic:2004ya}, it has also become possible to determine $V_{us}$ precisely by combining the experimental results for $K\to\pi$ semileptonic decays with lattice determinations of the form factor $f^+(0)=f^0(0)$, where the argument $0$ in the parentheses indicates that the four-momentum transfer $q$ between the kaon and pion satisfies $q^2=0$~\cite{Boyle:2007qe,Lubicz:2009ht,Boyle:2010bh}. The FLAG collaboration summarises the current status of the results for the form factor as~\cite{Colangelo:2010et}
\begin{equation}\label{eq:f0status}
f^+(0) = 0.956 \pm 0.008\,.
\end{equation}
For both the quantities $f_K/f_\pi$ and $f^+(0)$, the calculational techniques are such that we would obtain precisely 1 in the SU(3) flavour symmetry limit ($m_u=m_d=m_s$) and so it is the difference from one which we are actually computing. The Ademollo-Gatto theorem implies that these corrections are small for $f^+(0)$, and the main uncertainty for this quantity is due to the chiral extrapolation. The precision of the results in eqs.\,(\ref{eq:fkfpistatus}) and (\ref{eq:f0status}) is truly remarkable when compared to what was possible just a few years ago.

\begin{figure}[t]
\begin{center}
\includegraphics[width=.5\hsize]{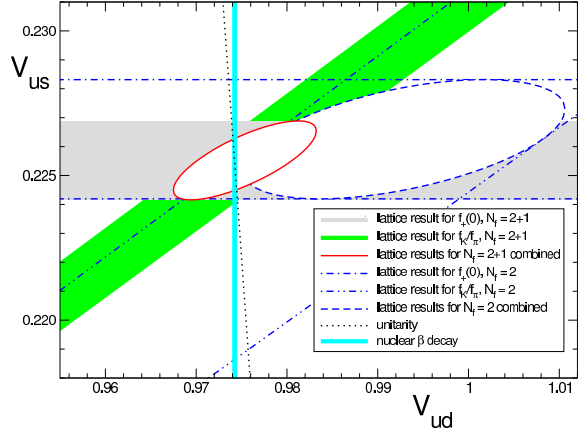}%
\end{center}\caption{The plot compares the information for $|V_{ud}|$ and $|V_{us}|$ obtained on the lattice with
the experimental result extracted from nuclear $\beta$ transitions. The dotted arc is part of the circle $|V_{ud}|^2+|V_{us}|^2=1$, representing the
correlation between $|V_{ud}|$ and $|V_{us}|$ that follows if the three-flavour CKM-matrix is unitary.~\cite{Colangelo:2010et}.\label{fig:vusflag}}
\end{figure}

The recent results are nicely summarized in figure~\ref{fig:vusflag}, also from FLAG~\cite{Colangelo:2010et}, in which are shown:
\begin{enumerate}
\item[(i)] ~the allowed regions from calculations of $f_K/f_\pi$ with $N_f=2+1$ and $N_f=2$ flavours of sea-quark (shown separately);
\item[(ii)] ~the allowed regions from calculations of $f^+(0)$ with $N_f=2+1$ and $N_f=2$ flavours of sea-quark (shown separately);
\item[(iii)] ~the allowed regions combining the calculations of $f_K/f_\pi$ and $f^+(0)$ with $N_f=2+1$ (red oval) and $N_f=2$ (dashed blue oval).
\end{enumerate}
Also shown on this plot is the result for $V_{ud}$ from super-allowed nuclear $\beta$-decays and an arc of the circle $|V_{ud}|^2+|V_{us}|^2=1$. Since $|V_{ub}|^2$ is very small compared to the uncertainties in $|V_{ud}|^2$ and $|V_{us}|^2$, this circle represents the unitarity condition on the first row of the CKM matrix. Although it would have been more exciting to find an inconsistency, we see that everything is remarkably consistent with the standard model, a point I underline further in the next subsection.

\subsection{$V_{us}$ within the Standard Model}
I'd now like to share a simple analysis, which I learned from my colleagues in the FLAG~\cite{Colangelo:2010et} collaboration,
which determines $V_{us}$ in the Standard Model without lattice calculations and which underlines again how remarkably consistent the lattice results for $f_K/f_\pi$ and $f^{+}(0)$ are with Standard Model expectations. Experimental studies of leptonic decays of kaons and pions give us two precise constraints for the four quantities, $V_{ud}$, $V_{us}$, $f_K/f_\pi$ and $f^{+}(0)$~\cite{Antonelli:2008jg}:
\begin{equation}\label{eq:kl2kl3}
\left|\frac{V_{us}\,f_K}{V_{ud}\,f_\pi}\right|=0.27599(59)\quad\textrm{and}\quad|V_{us}\,f^+(0)|=0.21661(47)\,.
\end{equation}
Within the standard model, a third equation is provided by the unitarity constraint:
\begin{equation}\label{eq:unitarity}
|V_{ud}|^2+|V_{us}|^2=1\,,
\end{equation}
where $|V_{ub}|^2$ has been dropped from the left-hand side because it is smaller than the uncertainties in the remaining terms. Now eqs.\,(\ref{eq:kl2kl3}) and (\ref{eq:unitarity}) give 3 equations for four unknowns. As the fourth equation we can take the value of $V_{ud}$ from the recent analysis in ref.\,\cite{Hardy:2008gy} based on 20 different superallowed nuclear $\beta$-transitions:
\begin{equation}\label{eq:vud}
\left|\,V_{ud}\,\right|=0.97425(22)\,.
\end{equation}
Combining equations (\ref{eq:kl2kl3}), (\ref{eq:unitarity}) and (\ref{eq:vud}) we obtain:
\begin{equation}\label{eq:threeresults}
|V_{us}|=0.22544(95),\quad f^+(0)=0.9608(46),\quad \frac{f_K}{f_\pi}=1.1927(59)\,,
\end{equation}
in excellent agreement with the lattice results for $f_K/f_\pi$ and $f^+(0)$ in (\ref{eq:fkfpistatus}) and (\ref{eq:f0status}).

\subsection{$K_{\ell 3}$ decays and SU(2) chiral perturbation theory}
\begin{table}[t]
\begin{center}
\begin{tabular}{cc||cc}\\
$m_\pi$&$f^{0}(q^2_{\rm max})$&$m_\pi$&$f^{0}(q^2_{\rm max})$\\ \hline
670\,MeV&1.00029(6)&575\,MeV&1.00016(6)\\
555\,MeV&1.00192(34)&470\,MeV&1.00272(34)\\
415\,MeV&1.00887(89)&435\,MeV&1.00416(43)\\
330\,MeV&1.02143(132)&375\,MeV&1.00961(123)\\
--&--&300\,MeV&1.01923(121)\\
--&--&260\,MeV&1.03097(224)\\
\end{tabular}
\caption{Values of $f^{0}(q^2_{\rm max})$ as a function of $m_\pi$ obtained by the RBC-UKQCD collaboration with 2+1 flavours of Domain Wall Fermions (left two columns)~\cite{Boyle:2007qe} and by the ETM collaboration with 2 flavours of twisted mass fermions (right two columns)~\cite{Lubicz:2009ht}.\label{tab:qsqmax}}
\end{center}
\end{table}

I end this section with some observations relating to semileptonic $K\to\pi$ decays ($K_{\ell 3}$ decays) using SU(2) chiral perturbation theory. Lattice calculations of $f^+(0)=f^{0}(0)$ start with the evaluation of $f^{0}(q^2_{\rm max})$, where $q^2_{\rm max}=(m_K-m_\pi)^2$ is the largest physical value of $q^2$ (where $q$ is the momentum transfer) and corresponds to the kinematics in which the initial kaon and final pion are both at rest. $f^{0}(q^2_{\rm max})$ can be calculated with excellent precision as illustrated in table\,\ref{tab:qsqmax}, where the results from the RBC-UKQCD and ETM collaborations are presented as functions of the pion mass. The reason that I present these results here, is that the entries in table~\ref{tab:qsqmax} appear to a long way below the value expected in the SU(2) chiral limit. This value is given by
one of the very few analytic non-perturbative results in QCD, the Callan-Treiman relation, which I present here in the form
\begin{equation}\label{eq:callantreiman}
f^{0}(q^2_{\rm max})=\frac{f_K}{f_\pi}\simeq 1.26\,
\end{equation}
where all the quantities are evaluated in the SU(2) chiral limit, $m_u=m_d=0$\,. It is a little puzzling that while the values of $f^{0}(q^2_{\rm max})$ in table~\ref{tab:qsqmax} are increasing as $m_\pi$ decreases, they are only doing so very slowly. In ref.~\cite{Flynn:2008tg}, we used SU(2) chiral perturbation theory to investigate whether the very slow increase in $f^{0}(q^2_{\rm max})$ observed in table\,\ref{tab:qsqmax} as $m_\pi$ decreases is expected to accelerate towards the value in eq.\,(\ref{eq:callantreiman}) for smaller values of the pion mass. We found that the one-loop chiral logarithms have a large coefficient and are of the correct size to account for the difference but they have the wrong sign, implying that the analytic terms (or higher order chiral logarithms) should account for approximately twice the difference between the results in table\,\ref{tab:qsqmax} and eq.\,(\ref{eq:callantreiman}). Of course the analytic terms (both linear and quadratic in $m_\pi$) are proportional to unknown low-energy constants (LECs) and hence are not calculable. In SU(3) chiral perturbation theory on the other hand, at one-loop order the LECs can be expressed in terms of $f_K/f_\pi$~\cite{Gasser:1984ux,Gasser:1984gg}. Estimating the SU(2) LECs by converting results from SU(3) ChPT suggests that the analytic terms have the correct sign and (large) magnitude to account for the difference.

It should be stressed here that since, apart from the values of the LECs, the above discussion relied only on SU(2) chiral perturbation theory and therefore the same features hold for $B\to\pi$ and $D\to\pi$ semileptonic decays.

Of course, we really want to know the chiral behaviour of the form-factor at the standard reference point $q^2=0$. At this point, the energy of the pion in the rest frame of the kaon is approximately $m_K/2$ and so cannot be considered soft in SU(2) chiral perturbation theory. In spite of the hard external pion, we found that it is possible to evaluate the chiral logarithms, since they can be calculated from soft \textit{internal} loops, finding~\cite{Flynn:2008tg}
\begin{eqnarray}
    f^{0}(0)=f^+(0)&=&F_+\left(1-\frac34 \frac{m_\pi^2}{16\pi^2f^2}\log\left(\frac{m_\pi^2}{\mu^2}\right)+c_+m_\pi^2\right)\label{eq:f0qsq0}\\
    f^-(0)&=&F_-\left(1-\frac34 \frac{m_\pi^2}{16\pi^2f^2}\log\left(\frac{m_\pi^2}{\mu^2}\right)+c_-m_\pi^2\right)\,,\label{eq:fminusqsq0}
\end{eqnarray}
where $F_{\pm}$ and $c_\pm$ are unknown LECs. In this way we have some information about the chiral behaviour of the form factors at $q^2=0$. Since the chiral extrapolation of the lattice results is the largest uncertainty in $f^0(0)$ any information about the chiral behaviour is useful (it would be very useful indeed if the results in eqs.(\ref{eq:f0qsq0}) and (\ref{eq:fminusqsq0}) could be extended to two loops in chiral perturbation theory). This \textit{hard-pion chiral perturbation theory} approach has been generalized to $K\to\pi\pi$ decays~\cite{Bijnens:2009yr} and to $B\to\pi$ and $D\to\pi$ semileptonic decays~\cite{Bijnens:2010ws}.

\section{$B_K$}\label{sec:bk}

\vspace{-.1in}One of the very important quantities in particle physics phenomenology for which the precision of lattice results has improved hugely in recent years has been $B_K$, the bag parameter of neutral kaon mixing. It is defined as the suitably normalized matrix element of the $\Delta S=2$ operator $(\bar d_L\gamma^\mu s_L)\,(\bar d_L\gamma^\mu s_L)$ between an initial $\bar{K}^0$ and final $K^0$ state (where $L$ denotes the left-handed component of the spinor field). As recently as five years ago, Chris Dawson, the rapporteur at this conference, was reporting results with a 12\% error largely due to the fact that there were insufficient unquenched results at light masses~\cite{Dawson:2005za}. A detailed review of recent results for $B_K$ has been presented at this conference by Jack Laiho~\cite{Laiho}, for illustration here I tabulate the key results in table~\ref{tab:compareres}. Not only are the errors reduced by a factor of 3 or so, but the results in table~\ref{tab:compareres} were obtained with a variety of fermion actions and techniques adding significantly to our confidence in the evaluation of the systematic uncertainties.

\begin{table}[t]\begin{center}
\begin{tabular}{ccc}\\
Publication &$N_f$ & $\hat{B}_K$\\ \hline\hline
\hline
RBC-UKQCD 2007\cite{Antonio:2007pb}  &2+1 & 0.720(13)(37)\\
Aubin et al. 2009   \cite{Aubin:2009jh} &2+1 & 0.724(8)(29)\\
Bae et al. 2010     \cite{Bae:2010ki}   & 2+1 & 0.724(12)(43)\\
RBC-UKQCD 2010  \cite{Aoki:2010pe}&2+1 & 0.749(7)(26)\\
\hline
JLQCD 2008  \cite{Aoki:2008ss}   & 2 & 0.758(6)(71)\\
ETMC 2010 \cite{Constantinou:2010qv}& 2& 0.729(30)\\
\end{tabular}\end{center}
\caption{A comparison of recent results for the renormalization group invariant $\hat{B}_K$. $N_f$ denotes the number of dynamical quark flavours. In each case, the first error is statistical and the second is systematic.
\label{tab:compareres}}
\end{table}

Lattice results for $B_K$ are an important ingredient in global studies of the unitarity triangle. Although in general the remarkable consistency of the information from different physical processes significantly restricts the possible parameter space for new physics, a number of \textit{tensions} have arisen in recent years at the 1.5\,--\,3 standard deviation level (see for example \cite{Charles:2004jd,Lunghi:2008aa,Bona:2009cj,Lenz:2010gu,Lunghi:2010gv}), and the results for $B_K$ contribute to these tensions. Calculations of $B_K$ will continue towards 1\% precision, but already with current precision it is necessary to begin considering corrections which had previously been neglected. For example it has been stressed that in the theoretical expression for $\epsilon_K$, the parameter monitoring indirect CP-violation in $K\to\pi\pi$ decays,  we should include terms proportional to Im\,$A_0$/Re\,$A_0$  (where $A_0$ is the $K\to\pi\pi$ amplitude with the two pions in a state with isospin 0) and recognise that the phase $\arctan (2\Delta M_K/\Delta\Gamma)$ is not precisely equal $\pi/4$ ($\Delta M_K$ and $\Delta\Gamma$ are the differences of the masses and widths of the $K_L$ and $K_S$ mesons)~\cite{Buras:2008nn,Buras:2009pj,Buras:2010pza}. Necessary theoretical improvements include the evaluation of long distance effects, for which we need to determine matrix elements of the time-ordered product of two operators (see sec.\,\ref{sec:long-distance}).

\section{$K\to\pi\pi$ decays}\label{sec:kpipi}
An important challenge for the lattice phenomenology community is the reliable calculation of $K\to\pi\pi$ decay amplitudes in general and attempts to reproduce the the experimental value of $\epsilon^\prime/\epsilon$ and to understand the $\Delta I=1/2$ rule (the enhancement of the $\Delta I=1/2$ amplitude by a factor of about 22 relative to that for $\Delta I=3/2$ transitions) in particular. The non-zero experimental value for $\epsilon'/\epsilon$ was historically the first evidence for direct CP-violation. In the past lattice calculations have tried to estimate the matrix elements by combining computations of $K\to\pi$ and $K\to$ vacuum matrix elements with the lowest order terms in the chiral expansion and I start with a brief review of the status of such calculations before proceeding to a discussion of the direct evaluation of the matrix elements with two-pion final states.

\subsection{$K\to\pi\pi$ decay amplitudes from $K\to\pi$ and $K\to$ vacuum matrix elements}\label{subsec:ktopi}
\begin{table}[t]
\begin{center}
\begin{tabular}{c|c|c}\hline \rule[-0.2cm]{0cm}{0.55cm}Collaboration(s)&Re
$A_0$/Re $A_2$ & $\varepsilon^\prime/\varepsilon$\\ \hline
\rule[-0.2cm]{0cm}{0.55cm}RBC~\cite{Blum:2001xb}& $25.3\pm 1.8$ &$-(4.0\pm 2.3)\times 10^{-4}$\\
\rule[-0.2cm]{0cm}{0.55cm}CP-PACS~\cite{Noaki:2001un}
& 9$\div$12&(-7\,$\div$\,-2)$\times 10^{-4}$\\
Experiments & 22.2 & $(17.2\pm 1.8)\times 10^{-4}$\\ \hline
\end{tabular}
\caption{\label{tab:kpiold} Quenched 2001 results on the $\Delta I=1/2$ rule and $\epsilon^\prime/\epsilon$ obtained from $K\to\pi$ and $K\to$ vacuum matrix elements using the lowest order term in the SU(3) chiral expansion.}
\end{center}\end{table}

\begin{figure}[t]
\begin{center}
\includegraphics[width=0.5\hsize]{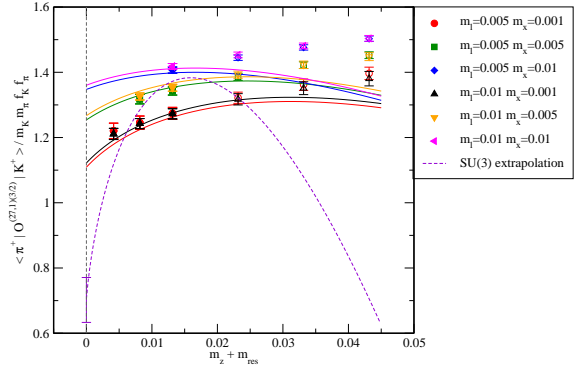}%
\end{center}
\caption{The mass dependence of the $K\to\pi$ matrix element. $m_l$, $m_x$ and $m_z$ are the masses of the sea and valence light quark and the valence "heavy" (strange) quarks respectively.\label{fig:kpiextrap}}
\end{figure}

At lowest order in the $SU(3)$ chiral expansion one can determine the $K\to\pi\pi$ decay amplitudes by calculating $K\to\pi$ and $K\to$\,vacuum matrix elements. In 2001, two collaborations published sone interesting quenched results on non-leptonic kaon decays in general and on the $\Delta I$=1/2 rule and $\epsilon^\prime/\epsilon$ in particular (see table~\ref{tab:kpiold}). In spite of the limitations of these calculations, the authors did achieve the control of the \textit{ultraviolet} problem, i.e. the numerical subtraction of power divergences and the renormalization of the weak operators. This is highly non-trivial.

The RBC-UKQCD collaboration have repeated the calculation in the pion-mass range 240-420\,MeV~\cite{Li:2008kc,Christ:2009ev}. For illustration consider the determination of $\alpha_{27}$, the lowest order LEC for the $\Delta I=3/2$ (27,1) operator:
\begin{equation}
O^{3/2}_{(27,1)}=(\bar{s}d)_L\,\big\{(\bar{u}u)_L-(\bar{d}d)_L\big\}+(\bar{s}u)_L\,(\bar{u}d)_L\,.
\end{equation}
Satisfactory fits for the mass dependence were obtained using NLO SU(3) ChPT, but the corrections were found to be very large, casting serious doubt on the approach. This is illustrated in fig.\ref{fig:kpiextrap} where the dashed curve represents the results in the SU(3) limit ($m_u=m_d=m_s$) and the value of this curve in the chiral limit is much below the data points, demonstrating that the one-loop corrections are very large. Thus the use of soft-pion theorems is not sufficiently reliable and $K\to\pi\pi$ matrix elements have to be computed directly. At this conference we have also heard about a proposed method to combine $K\to\pi$ matrix elements with $K\to\pi\pi$ ones limited to the two pions at rest~\cite{Laiho:2010ir}.

\subsection{Direct Calculations of $K\to\pi\pi$ Decay Amplitudes}\label{subsec:ktopipidirect}
From the above discussion we conclude that we must calculate $K\to\pi\pi$ matrix elements directly and the RBC-UKQCD collaboration is now undertaking a major study. The \textit{ultraviolet} problem of the subtraction of power divergences (i.e. of terms which diverge as inverse powers of the lattice spacing) remains tractable at the expense of significant statistical errors. The \textit{infrared} problem of extracting the spectrum and amplitudes in a finite Euclidean volume is also understood as long as we neglect the inelastic contribution (i.e. rescattering into states other than two pions). Assuming that the rescattering is dominated by the s-wave states, the quantization condition for two-pion states in a finite volume derived by L\"uscher takes the form~\cite{Luscher:1986pf,Luscher:1990ux,Luscher:1991cf}
\begin{equation}\label{eq:luscher}
\delta(q^\ast)+\phi(q^\ast)=n\pi\,
\end{equation}
where $\delta$ is the physical s-wave $\pi\pi$ phase shift, $\phi$ is a known kinematic function and $n$ is an integer. The relative momentum $q^\ast$ is related to the two-pion energy $E$ by $E^2=4(m_\pi^2+q^{\ast\,2})$. For illustration, all the formulae in this section are presented in the centre-of-mass frame, but they have also been generalised to moving frames~\cite{Rummukainen:1995vs,Kim:2005gf,Christ:2005gi}. Thus from the measured values of the energies one can determine the phase-shift. The relation between the measured Euclidean matrix elements and the physical amplitudes is given by~\cite{Lellouch:2000pv,Lin:2001ek}
\begin{equation}\label{eq:ll}
|A|^2=8\pi V^2\,\frac{m_K^3}{q^{\ast\,2}}\left\{\delta^\prime(q^\ast)+\phi^{P\,\prime}(q^\ast)\right\}\,|M|^2\,,
\end{equation}
where the $^\prime$ denotes differentiation w.r.t. $q^\ast$. (\ref{eq:ll}) has been generalized to moving frames in~\cite{Kim:2005gf,Christ:2005gi}.

\subsubsection{$K\to(\pi\pi)_{I=2}$ Decays}\label{subsubsec:ieq2}
The calculation of decay amplitudes into two pion states with isospin 2 is relatively straightforward; there are no power divergences to subtract nor any disconnected diagrams to evaluate. We will see in section\,\ref{subsubsec:ieq0} that these are significant difficulties when evaluating $\Delta I=1/2$ decay amplitudes. An exploratory quenched study with improved Wilson fermions was completed in 2004~\cite{Boucaud:2004aa}, but at that time we had not understood the finite-volume effects at non-zero total momentum. Results from RBC-UKQCD's exploratory quenched study with Domain Wall Fermions were presented in 2009~\cite{Lightman:2009cu} and this year M.Lightman presented results from dynamical simulations with almost physical pions, but on a course lattice~\cite{Goode:2011kb}. Before discussing the results I mention some theoretical points.

There are three $\Delta I=3/2$ operators whose matrix elements need to be evaluated:
\begin{eqnarray}
O^{3/2}_{(27,1)}&=&(\bar{s}^{\,i}d^i)_L\,\big\{(\bar{u}^ju^j)_L-(\bar{d}^jd^j)_L\big\}+(\bar{s}^{\,i}u^i)_L\,(\bar{u}^jd^j)_L \label{eq:o271def}\\
O^{3/2}_{7}&=&(\bar{s}^{\,i}d^i)_L\,\big\{(\bar{u}^ju^j)_R-(\bar{d}^jd^j)_R\big\}+(\bar{s}^{\,i}u^i)_L\,(\bar{u}^jd^j)_R \label{eq:o732def}\\
O^{3/2}_{8}&=&(\bar{s}^{\,i}d^j)_L\,\big\{(\bar{u}^ju^i)_R-(\bar{d}^jd^i)_R\big\}+(\bar{s}^{\,i}u^j)_L\,(\bar{u}^jd^i)_R\,,
\label{eq:o832def}
\end{eqnarray}
where $i$ and $j$ are colour indices and $(\bar{q}_1 q_2)_{L,R}=\bar{q}_1\gamma^\mu(1\mp\gamma^5)q_2$. The subscript $(27,1)$ in (\ref{eq:o271def}) denotes that the operator transforms as the $(27,1)$ representation of the $SU(3)_L\times SU(3)_R$ chiral symmetry group. $O_{7,8}^{3/2}$ are the $I=3/2$ components of the electroweak penguin operators labelled $O_{7,8}$ in the standard notation for the $\Delta S=1$ effective Hamiltonian and transform as the $(8,8)$ representation of $SU(3)_L\times SU(3)_R$.

A significant simplification in the calculation of the matrix elements of these operators is the use of the Wigner-Eckart theorem to relate the physical $K^+\to\pi^+\pi^0$ matrix elements to unphysical $K^+\to\pi^+\pi^+$ ones~\cite{Kim:2005gka}:
\begin{equation}\label{eq:wignereckart}
_{I=2}\langle\pi^+(p_1)\pi^0(p_2)\,|O^{3/2}_{1/2}|\,K^+\rangle=\frac32\langle \pi^+(p_1)\pi^+(p_2)\,|O^{3/2}_{3/2}|\,K^+\rangle\,,
\end{equation}
where the subscript on the operators indicates $\Delta I_z$, the change in the $z$-component of isospin. Eq.\,(\ref{eq:wignereckart}) is an exact relation in the isospin limit $m_u=m_d$, so that the evaluation of the matrix elements of the operators $O^{3/2}_{3/2}$ and $O^{3/2}_{1/2}$ are equivalent, but there are a number of advantages in using the fully extended $|\pi^+\pi^+\rangle$ states. The flavour structure of the operators $O^{3/2}_{3/2}$ is simply $(\bar{s}d)(\bar{u}d)$ rather than the $(\bar{s}d)(\bar{u}u-\bar{d}d)+(\bar{s}u){\bar{u}d}$ structure of the operators $O^{3/2}_{1/2}$ in eq.\,(\ref{eq:o271def})\,--\,(\ref{eq:o832def}).

With the $\pi^+\pi^+$-state we can impose antiperiodic boundary conditions on the $d$-quark say, so that the ground state is $|\pi^+(\pi/L)\pi^+(-\pi/L)\rangle$, where the arguments represent the momenta in lattice units (up to finite-volume corrections). In contrast to the use of periodic boundary conditions, for which the ground state corresponds to the two-pions at rest, it is not now necessary to isolate an excited state in order to have a decay into two-moving pions. For the physical decay, the minimum size of the lattice is halved from about 6\,fm to 3\,fm.

The final theoretical point I wish to make here is that the use of the Wigner-Eckart theorem also allows us evaluate the Lellouch-L\"uscher factor relating the measured matrix elements to the physical amplitudes directly~\cite{Kim:2010sd}. In particular, this factor requires knowledge of the derivative of the phase-shift  (see eq.\,(\ref{eq:ll})). By imposing partially-twisted boundary conditions~\cite{Bedaque:2004kc,deDivitiis:2004kq,Sachrajda:2004mi,Bedaque:2004ax} on the $d$-quark with twisting angle $\theta$ (it is sufficient to perform the twist in a single direction), the two-pion ground state now corresponds to a pion with momentum $\theta/L$ and the second pion with momentum $(\theta-2\pi)/L$. The corresponding $\pi\pi$ s-wave phase-shift can then be obtained by the L\"uscher formula\,(\ref{eq:luscher}) as a function of $\theta$ which allows for the derivative of the phase-shift to be evaluated directly at the masses being simulated. This procedure was tested in an exploratory simulation~\cite{Kim:2010sd} but has not yet been implemented in the main RBC-UKQCD programme.

\begin{figure}[t]
\begin{center}\begin{picture}(100,60)(-10,10)
\Line(0,10)(40,30)\Line(0,10)(80,10)
\Line(40,30)(80,10)\Oval(60,50)(5,28.28)(45)
\Text(-6,10)[r]{\scriptsize$K$}\Text(85,10)[l]{\scriptsize$\pi^+$}
\Text(85,70)[l]{\scriptsize$\pi^+$}\Text(36,32)[rb]{\scriptsize$O^{3/2}_{3/2}$}
\Text(20,23)[b]{\scriptsize{$s$}}
\CCirc(80,10){3}{Cyan}{Cyan}\CCirc(0,10){3}{Cyan}{Cyan}
\CCirc(80,70){3}{Cyan}{Cyan}\CCirc(40,30){5}{Red}{Yellow}
\end{picture}
\caption{Schematic diagram illustrating the contractions for $\Delta I=3/2$ $K\to\pi\pi$ decays in the RBC-UKQCD calculation~\cite{Lightman:2009cu}.\label{fig:deltai32diag}}
\end{center}\end{figure}
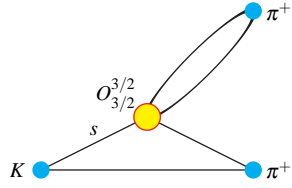

I now briefly summarise the results from the RBC-UKQCD Collaboration presented at this conference~\cite{Goode:2011kb}. The correlation functions which need to be evaluated are illustrated in fig.\ref{fig:deltai32diag}. The simulations were performed on $32^3\times 64\times 32$ lattice using a Domain Wall Fermion action for the quarks and the DSDR (dislocation suppressing determinant ratio) gauge action on a course lattice ($a^{-1}\simeq 1.4$\,GeV). The motivation for such a coarse lattice was to enable almost physical pions to be simulated, the $K\to\pi\pi$ amplitudes were obtained with a partially quenched pion of mass $m_\pi=145(6)$\,MeV (the corresponding unitary pion has $m_\pi\simeq 180$\,MeV) and the kaon mass is 519(2)\,MeV, both close to their physical values ($m_\pi=139.6$\,MeV, $m_K=493.7$\,MeV). The properties of these ensembles were discussed at this conference by Bob Mawhinney~\cite{boblatt2010}.

\begin{figure}[t]
\begin{center}
\includegraphics[width=0.31\hsize]{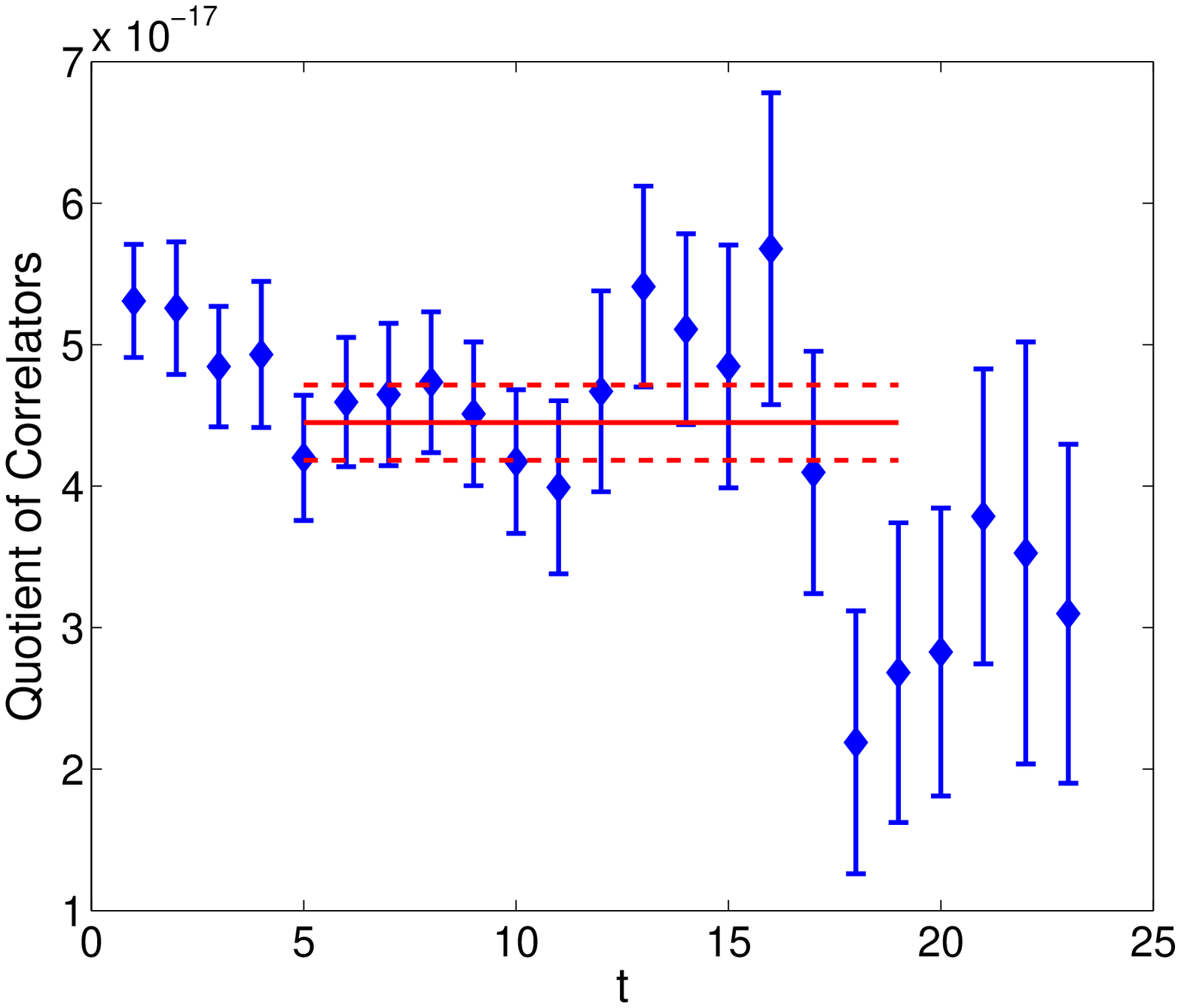}\quad
\includegraphics[width=0.31\hsize]{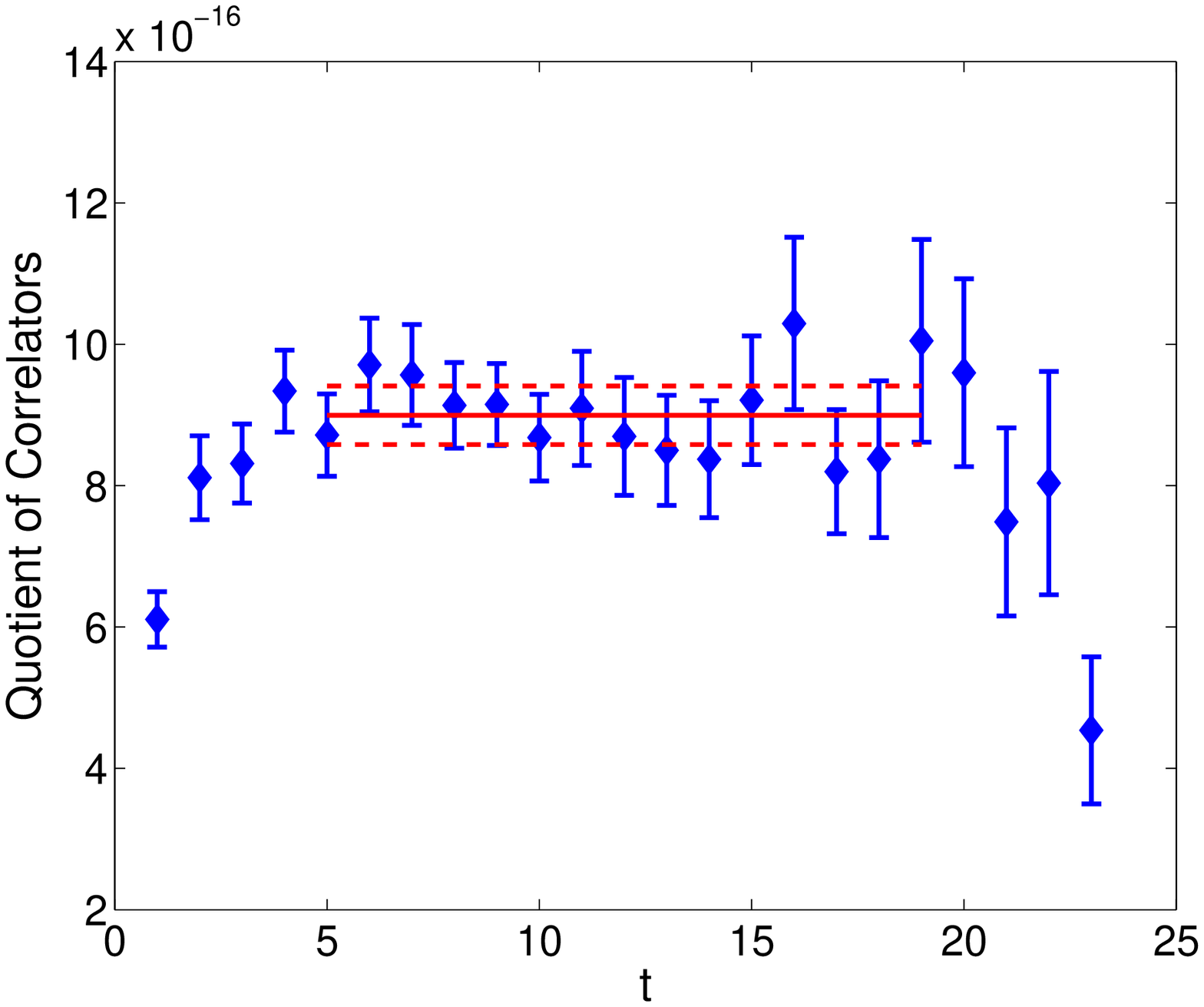}\quad
\includegraphics[width=0.31\hsize]{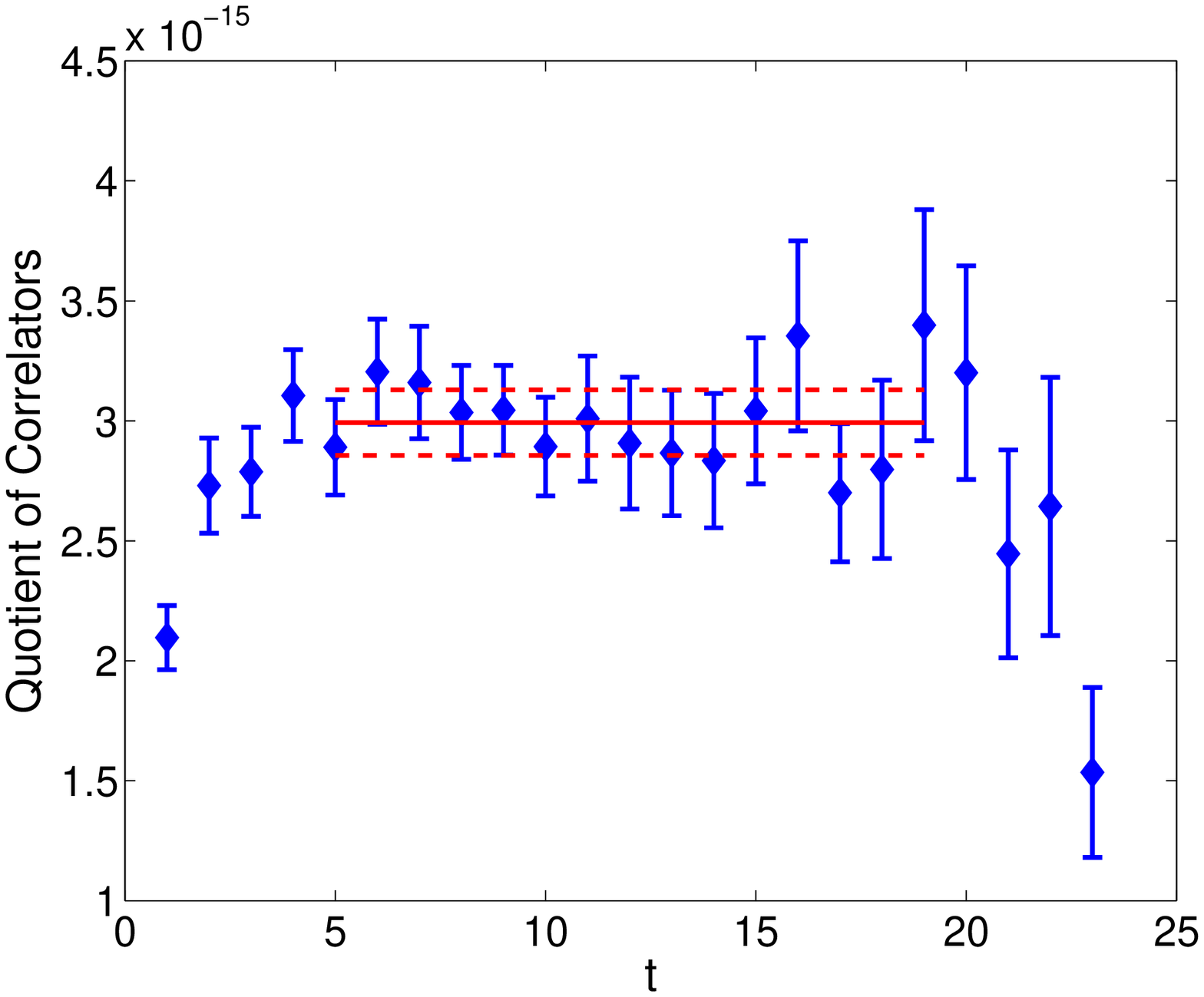}\quad
\end{center}
\vspace{-.2in}\hspace{0.925in}(a)\hspace{1.84in}(b)\hspace{1.835in}(c)
\caption{Sample plateau for the matrix elements (a) $(\bar{s}d)_L\,(\bar{u}d)_L$, (b) $(\bar{s}d)_L\,(\bar{u}d)_R$ and (c) $(\bar{s^i}d^j)_L\,(\bar{u^j}d^i)_R$, where $(\bar{s}d)_L\,(\bar{u}d)_{L,R}=(\bar{s^i}\gamma^\mu(1-\gamma^5) d^i)\,(\bar{u}^j\gamma_\mu (1\mp \gamma^5) d^j)$, and $i,j$ are colour labels.\label{fig:deltai32plateaus}}
\end{figure}

With the masses given above ($m_\pi=145.6(5)$\,MeV, $m_K=519(2)$\,MeV) and choosing the momentum of the pions to be $\sqrt{2}\,\pi/L$, the energy of the two-pion system is found to be 516(9)\,MeV so that the kinematics is almost matched. Sample plateaus for each of the three operators are presented in fig.\,\ref{fig:deltai32plateaus}. The preliminary result for the real part of the amplitude is Re\,$A_2=1.56(07)_{\rm stat}(25)_{\rm syst}\times 10^{-8}$\,GeV, to be compared to the experimental number of $1.50\times 10^{-8}$\,GeV\,. The non-perturbative renormalization has not been completed for the electroweak operators which contribute to Im\,($A_2$); guesstimating the renormalization factors RBC-UKQCD quote Im\,$(A_2)=-(9.6\pm0.04\pm 2.4)\times 10^{-13}$\,GeV~\cite{Goode:2011kb}, but after the renormalization is complete the systematic error will decrease from about 25\% to 15\%, dominated by the uncertainty in the value of the lattice spacing. It appears that these calculations are possible with good precision.

\subsubsection{$K\to(\pi\pi)_{I=0}$ Decays}\label{subsubsec:ieq0}

\begin{figure}[t]
\begin{center}
\includegraphics[width=0.7\hsize]{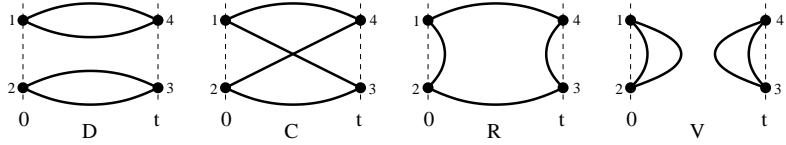}
\end{center}
\caption{The four diagrams which contribute to the two-pion propagator. \label{fig:twopiondiags}}
\end{figure}

The evaluation of $\Delta I=1/2$ matrix elements is very considerably harder than that for $\Delta I=3/2$ operators. The two-pion state with $I=0$ has vacuum quantum numbers and the vacuum contributions have to be subtracted requiring large statistical cancelations to obtain the $\exp(-E_{\pi\pi}\,t)$ behaviour, where $E_{\pi\pi}$ is the energy of the two-pion state. To illustrate this, consider the four diagrams in fig\,\ref{fig:twopiondiags} which make up the two-pion propagator. The $I=2$ $\pi\pi$ correlation function is proportional to D-C whereas the $I=0$ correlation function is proportional to 2D+C-6R+3V. The major practical difficulty is to subtract the vacuum contribution with sufficient precision.

\begin{figure}[t]
\begin{center}
\includegraphics[width=0.5\hsize]{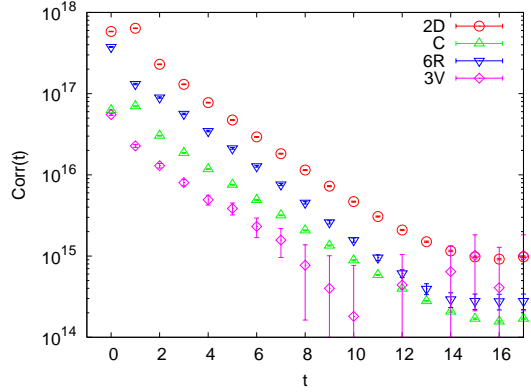}
\end{center}
\caption{Contribution of the four components to the $I=0$ two-pion correlation function.\label{fig:twopioncorr}}
\end{figure}

To demonstrate the above, I now present some results from the exploratory study by RBC-UKQCD on a $16^3\times 32\times 16$ lattice with an inverse lattice spacing of $a^{-1}=1.73$\,GeV and an unphysically heavy pion with mass 420\,MeV~\cite{Liu:2010fb}. To increase the statistical precision, quark propagators are generated from each of the 32 time slices. The four components of the $I=0$ two-pion correlation function 2D+C-6R+3V are shown separately in fig.\,\ref{fig:twopioncorr}~\cite{Liu:2010fb}, from which we see that the error on the 3V component grows significantly at larger times.

\begin{figure}
\begin{center}\begin{picture}(320,135)(-5,35)
\Line(0,140)(30,160)\Line(0,140)(60,130)\Line(30,160)(60,130)
\Oval(45,160)(5,15)(0)
\CCirc(60,160){1.5}{Blue}{Blue}\CCirc(60,130){1.5}{Blue}{Blue}
\CCirc(0,140){1.5}{Blue}{Blue}
\CCirc(30,160){3}{Cyan}{Cyan}
\Text(-4,140)[r]{$K$}\Text(64,160)[l]{$\pi$}\Text(64,130)[l]{$\pi$}
\Text(30,120)[t]{Type1}\Text(15,155)[b]{$s$}
\Oval(135,145)(5,15)(0)\CArc(210,145)(33.54,153.43,206.57)
\Line(150,145)(180,160)\Line(150,145)(180,130)
\CCirc(180,160){1.5}{Blue}{Blue}\CCirc(180,130){1.5}{Blue}{Blue}
\CCirc(120,145){1.5}{Blue}{Blue}
\CCirc(150,145){3}{Cyan}{Cyan}
\Text(116,145)[r]{$K$}\Text(184,160)[l]{$\pi$}\Text(184,130)[l]{$\pi$}
\Text(150,120)[t]{Type2}\Text(135,153)[b]{$s$}
\Line(270,152.5)(300,160)\Line(240,145)(270,152.5)
\Line(240,145)(300,130)\CArc(330,145)(33.54,153.43,206.57)
\Oval(270,162.5)(10,4)(0)
\CCirc(300,160){1.5}{Blue}{Blue}\CCirc(300,130){1.5}{Blue}{Blue}
\CCirc(240,145){1.5}{Blue}{Blue}
\CCirc(270,152.5){3}{Cyan}{Cyan}
\Text(236,145)[r]{$K$}\Text(304,160)[l]{$\pi$}\Text(304,130)[l]{$\pi$}
\Text(270,120)[t]{Type3}\Text(255,153)[b]{$s$}\Text(270,175)[b]{$l,s$}
\Oval(15,60)(7,15)(0)\GCirc(37.5,60){7.5}{1}
\CArc(90,60)(33.54,153.43,206.57)\CArc(30,60)(33.54,333.43,26.57)
\CCirc(60,75){1.5}{Blue}{Blue}\CCirc(60,45){1.5}{Blue}{Blue}
\CCirc(0,60){1.5}{Blue}{Blue}
\CCirc(30,60){3}{Cyan}{Cyan}
\Text(-4,60)[r]{$K$}\Text(64,75)[l]{$\pi$}\Text(64,45)[l]{$\pi$}
\Text(30,40)[t]{Type4}\Text(15,70)[b]{$s$}\Text(38,69)[b]{$l,s$}
\Line(150,72.5)(180,80)\Line(120,65)(150,72.5)
\Line(120,65)(180,50)\CArc(210,65)(33.54,153.43,206.57)
%\Oval(150,82.5)(10,4)(0)
\CCirc(180,80){1.5}{Blue}{Blue}\CCirc(180,50){1.5}{Blue}{Blue}
\CCirc(120,65){1.5}{Blue}{Blue}
%\CCirc(150,72.5){3}{Cyan}{Cyan}
\CBoxc(150,72.5)(5,5){Cyan}{Cyan}
\Text(116,65)[r]{$K$}\Text(184,80)[l]{$\pi$}\Text(184,50)[l]{$\pi$}
\Text(150,40)[t]{Mix3}\Text(135,73)[b]{$s$}
\Oval(255,60)(7,15)(0)
\CArc(330,60)(33.54,153.43,206.57)\CArc(270,60)(33.54,333.43,26.57)
\CCirc(300,75){1.5}{Blue}{Blue}\CCirc(300,45){1.5}{Blue}{Blue}
\CCirc(240,60){1.5}{Blue}{Blue}
\CBoxc(270,60)(5,5){Cyan}{Cyan}
\Text(236,60)[r]{$K$}\Text(304,75)[l]{$\pi$}\Text(304,45)[l]{$\pi$}
\Text(270,40)[t]{Mix4}\Text(255,70)[b]{$s$}
\end{picture}\end{center}
\caption{Types of correlation function corresponding to $K\to\pi\pi$ decays. The blue circles represent the insertion of a four quark operator appearing in the weak Hamiltonian and the squares that of the pseudoscalar density.\label{fig:kpipidiags}}
\end{figure}
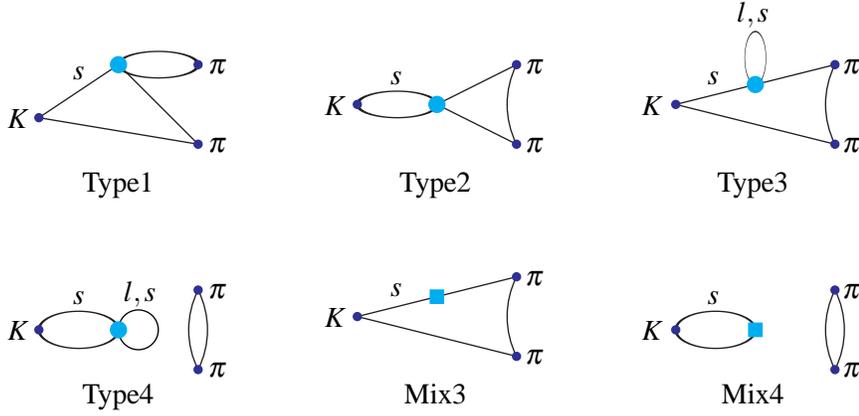

In spite of the difficulty of subtracting the vacuum contributions, at this conference Qi Liu presented the results of a complete calculation of $A_0$ and $A_2$, albeit at unphysical kinematics with $m_\pi\simeq 420$\,MeV and with the pions at rest~\cite{Liu:2010fb}. There are 6 diagrams, shown in fig.\,\ref{fig:kpipidiags} and a total of 48 Wick contractions. The diagrams labeled by Mix3 and Mix4 in fig.\ref{fig:kpipidiags} are needed to subtract the power divergences which are proportional to matrix elements of the pseudoscalar density $\bar{s}\gamma_5 d$. The results for this unphysical kinematics are~\cite{Liu:2010fb}
\begin{eqnarray}
\textrm{Re}~A_0=(3.0\pm 0.9)\,10^{-7}\,\textrm{GeV},&\quad&
\textrm{Im}~A_0=-(2.9\pm 2.2)\,10^{-11}\,\textrm{GeV},\\
\textrm{Re}~A_2=(5.394\pm 0.045)\,10^{-8}\,\textrm{GeV},&\quad&
\textrm{Im}~A_2=-(7.79\pm 0.08)\,10^{-13}\,\textrm{GeV}\,.
\end{eqnarray}
The precision is limited by the disconnected diagrams and the vacuum subtraction.
At least the calculation of the real part of $A_0$ appears to be tractable for this simplified kinematics at heavy pion masses, whereas for Im\,$A_0$ we will require more statistics to confirm that there is indeed a signal. The greater challenge is now to proceed towards performing the corresponding calculations at physical kinematics.

\section{$\eta$ and $\eta^\prime$ Mesons}\label{sec:etaetap}

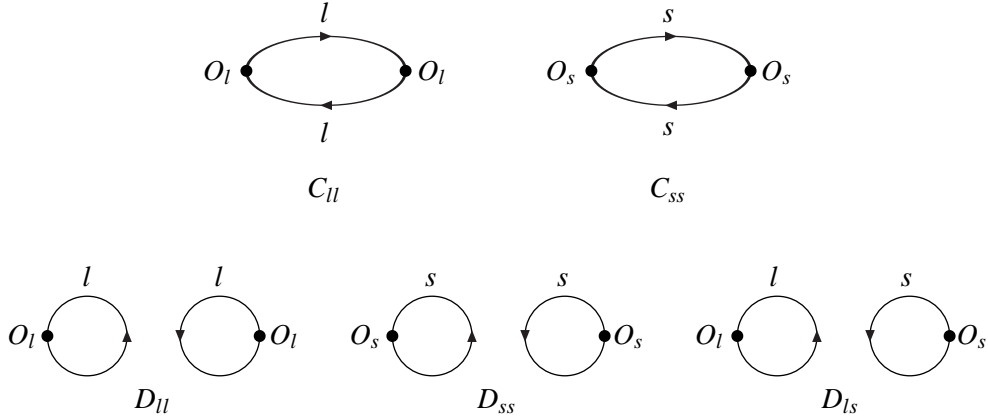
\begin{figure}[t]
\begin{center}
\begin{picture}(370,150)(0,20)
\GCirc(30,50){15}{1}\GCirc(15,50){2}{0}\ArrowLine(45,49.5)(45,50.5)
\Text(30,69)[b]{$l$}\Text(11,50)[r]{$O_l$}
\GCirc(80,50){15}{1}\GCirc(95,50){2}{0}\ArrowLine(65,50.5)(65,49.5)
\Text(80,69)[b]{$l$}\Text(99,50)[l]{$O_l$}
\Text(55,30)[t]{$D_{ll}$}
\GCirc(160,50){15}{1}\GCirc(145,50){2}{0}\ArrowLine(175,49.5)(175,50.5)
\Text(160,69)[b]{$s$}\Text(141,50)[r]{$O_s$}
\GCirc(210,50){15}{1}\GCirc(225,50){2}{0}\ArrowLine(195,50.5)(195,49.5)
\Text(210,69)[b]{$s$}\Text(229,50)[l]{$O_s$}
\Text(185,30)[t]{$D_{ss}$}
\GCirc(290,50){15}{1}\GCirc(275,50){2}{0}\ArrowLine(305,49.5)(305,50.5)
\Text(290,69)[b]{$l$}\Text(271,50)[r]{$O_l$}
\GCirc(340,50){15}{1}\GCirc(355,50){2}{0}\ArrowLine(325,50.5)(325,49.5)
\Text(340,69)[b]{$s$}\Text(359,50)[l]{$O_s$}
\Text(315,30)[t]{$D_{ls}$}
\Oval(120,150)(13,30)(0)\ArrowLine(119.5,163)(120.5,163)\ArrowLine(120.5,137)(119.5,137)
\GCirc(90,150){2}{0}\GCirc(150,150){2}{0}
\Text(120,169)[b]{$l$}\Text(85,150)[r]{$O_l$}
\Text(120,131)[t]{$l$}\Text(155,150)[l]{$O_l$}
\Text(120,110)[t]{$C_{ll}$}
\Oval(250,150)(13,30)(0)\ArrowLine(249.5,163)(250.5,163)\ArrowLine(250.5,137)(249.5,137)
\GCirc(220,150){2}{0}\GCirc(280,150){2}{0}
\Text(250,169)[b]{$s$}\Text(215,150)[r]{$O_s$}
\Text(250,131)[t]{$s$}\Text(285,150)[l]{$O_s$}
\Text(250,110)[t]{$C_{ss}$}
\end{picture}\end{center}
\caption{\label{fig:etas} Diagrams contributing to the correlation functions for the $\eta$\,-\,$\eta^\prime$ system. $l$ and $s$ represent the light ($u$ and $d$) and strange quarks respectively. The connected diagrams are denoted by $C_{ll}$ and $C_{ss}$ and the disconnected ones by $D_{ll}$, $D_{ss}$, $D_{ls}$ and $D_{sl}$ (not shown).}
\end{figure}

\begin{figure}[t]
\begin{center}
\includegraphics[width=0.4\hsize]{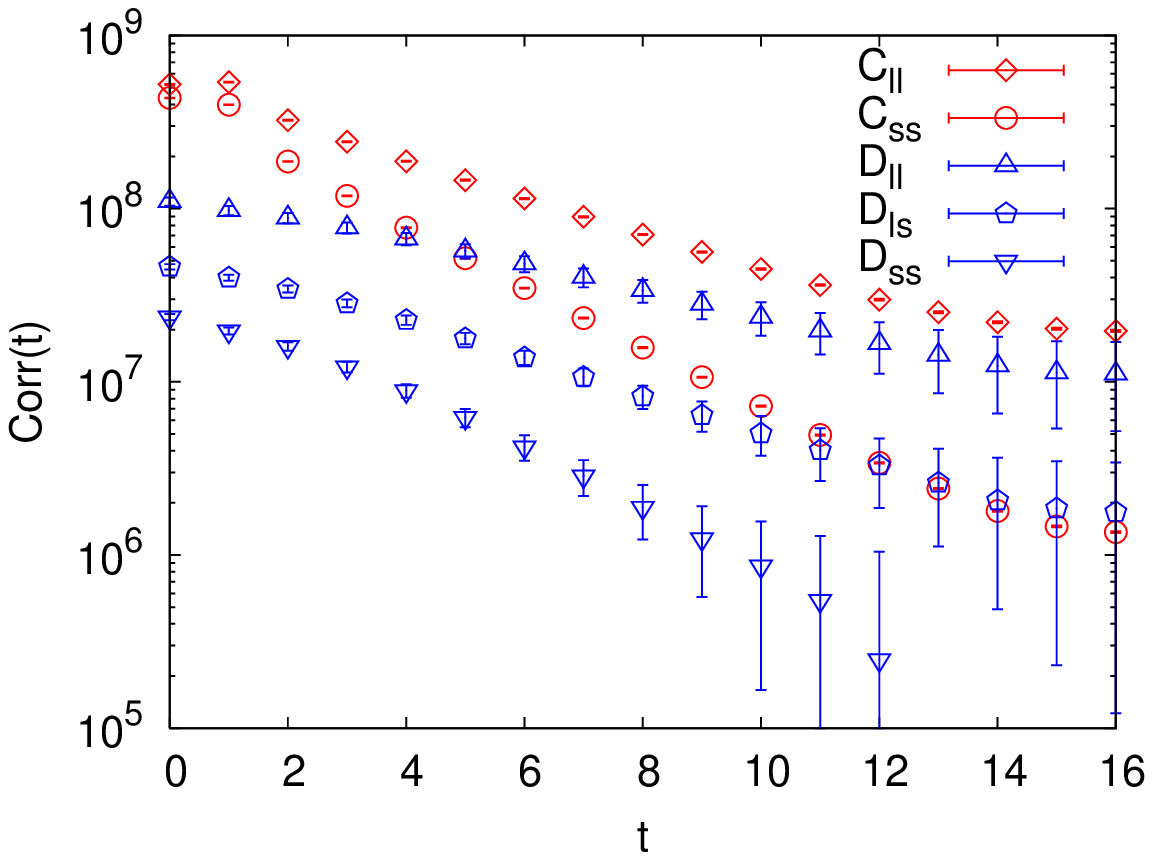}\hspace{0.5in}
\includegraphics[width=0.4\hsize]{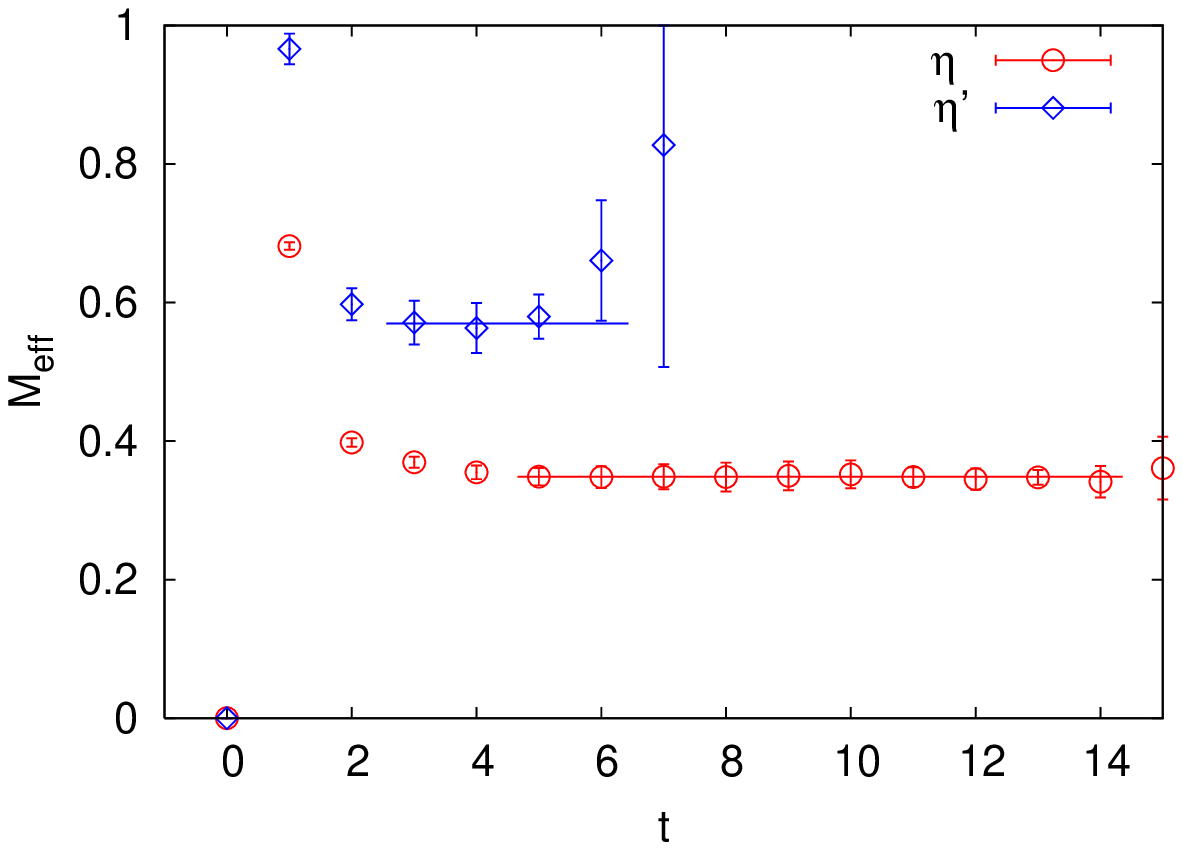}
\begin{picture}(300,20)(0,0)
\Text(56,10)[c]{(a)}\Text(264,10)[c]{(b)}\end{picture}
\end{center}\vspace{-.25in}\caption{\label{fig:etaplots}(a) Values of each of diagrams as a function of $t$. (b) The effective mass plots for the $\eta$ and $\eta^\prime$ mesons. In each case the results are for the lightest mass in the simulations; see \cite{Christ:2010dd} for details.}
\end{figure}

In the previous section we saw the importance of controlling disconnected diagrams and this is the case for many important phenomenological quantities. Here I mention one other longstanding issue, the spectrum and mixing of $\eta$ and $\eta^\prime$ mesons. To determine the propagators we need to evaluate the diagrams shown in fig.\,\ref{fig:etas} and I report here on our recent study on a lattice of spatial extent $16^3$ and lattice spacing $a^{-1}=1.73$\,GeV~\cite{Christ:2010dd}. Using interpolating operators of the form
\begin{equation}
O_l=\frac{\bar{u}\gamma_5 u+\bar{d}\gamma_5d}{\sqrt{2}}\quad\textrm{and}
\quad O_s=\bar{s}\gamma_5 s\,,
\end{equation}
we calculate the correlation functions
\begin{eqnarray}
X_{\alpha\beta}(t)&=&\frac{1}{32}\,\sum_{t^\prime=0}^{31}\ \langle\ O_\alpha(t+t^\prime)\,O_\beta(t^\prime)\ \rangle\quad\textrm{where}\quad \alpha,\beta=l,s\\
&=&\begin{pmatrix}C_{ll}-2D_{ll}&-\sqrt{2}D_{ls}\\ -\sqrt{2}D_{sl}&C_{ss}-D_{ss}\end{pmatrix}~,
\end{eqnarray}
where the diagrams are defined in fig.\,\ref{fig:etas}. The propagators are generated with sources at each of the 32 time slices $t^\prime$. In fig.\,\ref{fig:etaplots}(a) we present the values of the diagrams at the lightest mass used in the simulations as a function of $t$, from which we see that the usual expectation that disconnected diagrams are small does not apply here and also the expected feature that the uncertainties on the disconnected diagrams are significantly larger than those on the connected ones. The effective mass plots for the eigenvalues of $X(t)$ are shown in fig.\,\ref{fig:etaplots}(b), again at the lightest masses used in the simulations, from which we see a good plateau for the $\eta$ and a short, but clear, plateau for the $\eta^\prime$. After the chiral extrapolation we obtain the values $m_\eta=573(6)\,$MeV and $m_{\eta^\prime}=947(142)$\,MeV, where only the statistical errors are shown, compared to the experimental values of 548\,MeV and 958\,MeV respectively.

As explained in ref.\,\cite{Christ:2010dd}, it is also possible to determine the mixing angle. Once SU(3) flavour symmetry is broken it is not clear that the physical $\eta$ and $\eta^\prime$ states are simply linear combinations of the octet and singlet flavour combinations; nevertheless it is a standard phenomenological assumption that they are. Based on this assumption we obtain a value of $\theta=-14.1(2.8)^\circ$ (statistical error only) in agreement with the phenomenological estimates lying in the range $-20^\circ$ to $-10^\circ$. A particularly important feature of the calculation  is that it is possible to check the orthogonality of the mixing matrix confirming that it is a good approximation to consider only the symmetric octet and singlet states in order to understand the mixing.

In spite of the limited precision, our calculation demonstrates that QCD can explain the large mass of the ninth pseudoscalar meson and its small mixing with the SU(3) octet state and provides a first benchmark for future calculations. There remains much to be done now to reduce the statistical errors and to quantify in detail the systematic uncertainties. Progress in the efficient determination of all-to-all propagators will be very important in improving the reach and precision of lattice calculations of physical quantities in which disconnected diagrams play an major role.

\section{Long-Distance Contributions to Physical Quantities}\label{sec:long-distance}

We are used to calculating the short-distance contributions to physical processes, generally formulating the calculation as the evaluation of the matrix element of a local operator. A good example is the evaluation of the $B_K$ parameter in neutral-kaon mixing illustrated by
\begin{center}
\begin{picture}(240,50)(-10,0)
\Line(0,10)(60,10)\Line(0,40)(60,40)
\Photon(20,10)(20,40){2}{5}\Photon(40,10)(40,40){2}{5}
\Text(-4,10)[r]{$\overline{d}$}\Text(-4,40)[r]{$s$}
\Text(64,10)[l]{$\overline{s}$}\Text(64,40)[l]{$d$}
\Text(15,25)[r]{\scriptsize$W$}\Text(45,25)[l]{\scriptsize$W$}
\Text(90,25)[c]{\Large{=}}\Text(115,25)[l]{\large$C(M_W/\mu)$}
\Line(180,10)(200,25)\Line(180,40)(200,25)
\Line(220,10)(200,25)\Line(220,40)(200,25)
\CCirc(200,25){2}{Cyan}{Cyan}
\Text(176,10)[r]{$\overline{d}$}\Text(176,40)[r]{$s$}
\Text(224,10)[l]{$\overline{s}$}\Text(224,40)[l]{$d$}\,.
\end{picture}
\end{center}
The determination of the matrix element of the resulting $\Delta S=2$ local operator has been discussed in the section\,\ref{sec:bk}. In many cases the short-distance contribution is the dominant term, but long-distance contributions are not always negligible, for example if the GIM suppression is logarithmic or if there is a CKM enhancement (even if the GIM suppression is power like). As lattice results become more precise, we should try to compute the long-distance contributions effectively and this represents a new type of calculation. Early thoughts in this direction include ref.\,\cite{Isidori:2005tv} for rare kaon decays and ref.\cite{Christ:2010zz} for neutral kaon mixing.
\subsection{Rare Kaon Decays}\label{subsec:rarekaon}
In ref.\,\cite{Isidori:2005tv}, the authors conclude that it is possible in principle to evaluate the long distance effects for $K\to\pi\ell^+\ell^-$ and $K\to\pi\nu\bar\nu$ decays. This requires the evaluation of $T$-products of the form
\begin{equation}\label{eq:rarekaont}
T_{Q,J}(q^2)=N_V\int\,d^4x~d^4y~e^{-iq.y}~\langle\pi\,|\,T\left[Q(x)\,J^\mu(y)\right]\,|\,K\rangle
\end{equation}
where $Q$ is one of the four-quark operators appearing in the weak Hamiltonian, $J$ is a weak or electromagnetic current and $N_V$ is a volume factor. The generic lattice calculation is therefore of correlation functions of the form
\begin{equation}
-i\int\,d^4x~e^{-iq\cdot x}\langle0\,|\phi_\pi(t_\pi,\vec{p})\,J^\mu_X(x)\,\left[Q_i^u(0)-Q_i^c(0)\right] \,\phi_K^\dagger(t_K,\vec{k})\, |\,0\rangle\,,
\end{equation}
with $t_\pi>0$ and $t_K<0$. $\phi_\pi$ and $\phi_K$ are the interpolating operators for the $\pi$ and $K$ mesons respectively (after the three-dimensional Fourier transform has been taken). The main issue discussed in ref.\,\cite{Isidori:2005tv} is that of renormalization, the subtraction of power divergences and the consequences of contact terms. The authors conclude that, as a result of symmetries and the GIM mechanism, the power divergences can be removed and they check their general arguments by one-loop perturbative calculations. They believe that their study opens
a new field of interesting physical applications for the lattice community, although to date no such numerical calculations have been performed.

\subsection{Long-distance contribution to the $K_L$-$K_S$ mass difference and $\epsilon_K$}\label{subsec:klks}
At this conference Norman Christ presented some interesting ideas how the long-distance contribution to the $K_L-K_S$ mass difference and $\epsilon_K$ might be evaluated~\cite{Christ:2010zz}. This requires
\begin{enumerate}
\item[(i)]~the calculation of the long-distance contribution to the matrix elements of the product of two $\Delta S=1$, four-quark weak operators between kaon states;
\item[(ii)]~the subtraction of the short distance part of this matrix element in a way that is consistent with the original explicit evaluation of the short-distance contribution;
\item[(iii)]~a generalization of the Lellouch-L\"uscher approach to finite-volume corrections to second order in the weak interaction.
\end{enumerate}
Although much work remains to be done to develop these ideas into a practicable method, the main theoretical steps have now been taken.

\section{Non-leptonic $B$-Decays}\label{sec:nonleptB}
I end this talk with a discussion of an important class of processes in flavour physics for which, up to now at least, little or no progress has been made in formulating a lattice approach, namely non-leptonic $B$-decays. A huge amount of precise information, from over 100 channels, has been obtained about decay rates and CP-asymmetries for the exclusive decays of $B$-mesons into two light mesons. Unfortunately, with just a few exceptions (most notably the CP-asymmetry in the golden-mode $B\to J\psi K_S$), our ability to deduce fundamental information about CKM matrix elements is limited by our inability to quantify the non-perturbative QCD effects sufficiently precisely.

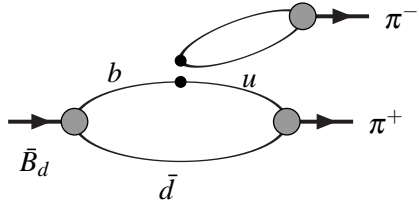
\begin{figure}[t]
\begin{center}\begin{picture}(140,75)(-65,-25)
\Oval(0,0)(15,40)(0)\Oval(23,31.37)(7,24.48)(20)
\SetWidth{1.5}
\ArrowLine(-65,0)(-40,0)
\ArrowLine(45,0)(65,0)\ArrowLine(51,39.74)(71,39.74)
\SetWidth{0.5}
\GCirc(-40,0){5}{0.6}\GCirc(40,0){5}{0.6}
\GCirc(46,39.74){5}{0.6}
\GCirc(0,15){2}{0}\GCirc(0,23){2}{0}
\Text(-55,-10)[t]{$\bar
B_d$}
\Text(77,39.74)[l]{$\pi^-$} \Text(71,0)[l]{$\pi^+$}
\Text(-25,15)[b]{$b$}\Text(26,13)[b]{$u$} \Text(-5,-20)[t]{$\bar
d$}
\end{picture}\end{center}
\caption{\label{fig:naive}Illustration of na\"ve factorization for the decay $\bar{B}_d\to\pi^+\pi^-$. The local four-quark operator $(\bar{u}b)_{V-A}\,(\bar{d}u)_{V-A}$ is shown as a product of the bilinear operators
$(\bar{u}b)_{V-A}$ and $(\bar{d}u)_{V-A}$ represented by the black circles. The two black circles are separated for clarity.}
\end{figure}

The operator product expansion can be used to separate the long and short distance physics, so that the non-perturbative QCD effects are contained in the matrix elements of local operators, $\langle M_1M_2|O_i(0)|B\rangle$, where $O_i$ is a $\Delta B=1$ operator and $M_{1,2}$ are light mesons. In contrast to $K\to\pi\pi$ decays discussed in section~\ref{sec:kpipi}, where it is a good approximation to restrict consideration of final-state rescattering to two-pion states, the large mass of the $B$-meson means that very many intermediate states contribute. We do not yet(?) have a theoretical framework for treating this in simulations in Euclidean space.

In the past phenomenological approaches to nonleptonic $B$-decays were based on na\"ive factorization in which the matrix element was reduced to products of matrix elements of $B\to M_1$ and vacuum$\to M_2$ and/or vice-versa, depending on the flavour quantum numbers. This is illustrated in fig.\,\ref{fig:naive}, where the matrix element of the local four-quark operator $(\bar{u}b)_{V-A}\,(\bar{d}u)_{V-A}$ is shown as the product of the matrix elements of the two currents:
\begin{equation}\label{eq:naive}
\langle\,\pi^+\pi^-\,|\,(\bar u b)_{V-A}\ (\bar d
u)_{V-A}\,|\,\bar B_d\,\rangle \overset{?}{=} \langle\,\pi^-\,|\,(\bar d
u)_{V-A}\,|\,0\,\rangle \ \langle\,\pi^+\,|\,(\bar u
b)_{V-A}\,|\,\bar B_d\,\rangle\,.
\end{equation}
The motivation for assuming (\ref{eq:naive}) is that the two factors on the right-hand side are known or calculable (in principle at least). The first factor is proportional to the leptonic decay constant $f_\pi$ and the second is proportional to the form-factors of semileptonic $B\to\pi$ decays. On the other hand the limitations of such a model are clear; the renormalization-scale dependence of the two sides do not match and rescattering effects are not included.

In 1999, together with Beneke, Buchalla and Neubert, we realised that as $m_b\to\infty$, the long-distance QCD effects do factorise into simpler universal quantities~\cite{Beneke:1999br,Beneke:2000ry,Beneke:2001ev}
\begin{eqnarray}
\langle\,M_1,M_2\,|\,O_i\,|\,B\,\rangle&=& \sum_jF_j^{B\to
M_1}(m_2^2)\int_0^1 du\,T_{ij}^I(u)\Phi_{M_2}(u)
+\ \ (M_1\,\leftrightarrow\,M_2)\nonumber\\
&&\hspace{-1in}+\int_0^1\,d\xi\,du\,dv\,T^{II}_i(\xi,u,v)\,\Phi_B(\xi)\,
\Phi_{M_1}(v)\,\Phi_{M_2}(u)\,,\label{eq:qcdfactorization}
\end{eqnarray}
where $F_j^{B\to M_1}$ are the form factors for $B\to M_1$ transitions, the $\Phi$ are light-cone distribution amplitudes and $T^{I,II}$ are short distance contributions and are calculable in perturbation theory. $u$ and $v$ are the momentum fractions carried by the quarks in the mesons. The significance of the factorization
formula stems from the fact that the non-perturbative quantities which appear on the right-hand side are much simpler than the original matrix elements. They either reflect universal properties of a single meson state
(the light-cone distribution amplitudes) or refer to a $B\to$ meson transition matrix element of a local current (form factors). Conventional (na\"ive) factorization is recovered as a rigorous prediction in the infinite mass
limit (i.e. neglecting $O(\alpha_s)$ and $O(\Lambda_{\textrm{{\tiny QCD}}}/m_b)$ corrections).
Perturbative corrections to na\"ive factorization can be computed systematically and the results are, in general,
process dependent. It is a remarkable feature that all strong interaction phases are
generated perturbatively in the heavy quark limit. The factorization formulae are valid up to
$O(\Lambda_{\textrm{\tiny{QCD}}}/m_b)$ corrections and the main limitation of the framework is due to the fact that since $m_b$ is not so large, CKM and chiral enhancements to non-factorizable $O(\Lambda_{\textrm{QCD}}/m_b)$ terms are important.

Although we do not know how to evaluate the matrix elements for $B\to M_1M_2$ decays directly, we can ask what can lattice simulations contribute to the factorization formula (\ref{eq:qcdfactorization})? The moments of the light-cone distribution functions for the light mesons $M_{1,2}$ can and are being computed~\cite{Arthur:2010xf} as are the $B\to M$ form factors. What we do not know how to compute at this stage are the parton distribution amplitudes of $\Phi_B$ or its moments and I end this section with a brief explanation of the reasons for this. $\Phi_B$ is defined by
\begin{equation}\label{eq:phiBdef}
\Phi_{B\,\alpha\beta}(\tilde{k}_+)=\int dz_-\,e^{i\tilde{k}_+z_-}\,\langle\,0\,|\,\bar{u}_\beta(z)[z,0]b_\alpha(0)\,|\,B\,\rangle\big |_{z_+,z_\perp=0}\,,
\end{equation}
where $\pm$ denote light-cone coordinate and $[z,0]$ represents the part-ordered exponential of gauge fields between $z$ and $0$. In evaluating matrix elements, $\Phi_B$ is convoluted with the perturbative hard-scattering amplitude $T_i^{II}$ and the relevant quantity is
\begin{equation}\label{eq:lambdab}
\frac{\sqrt{2}}{\lambda_b}=\int_0^\infty\,\frac{d\tilde{k}_+}{\tilde{k}_+}\,\Phi_B(\tilde{k}_+)\,.
\end{equation}
(In higher orders of perturbation theory factors containing $\log(\tilde{k}_+)$ appear.)
Although at large $\tilde{k}_+$, $\phi_B(\tilde{k}_+)\sim 1/\tilde{k}_+$, the convolution in eq.(\ref{eq:lambdab}) is finite. In lattice calculations, at least up to now, we know how to calculate the matrix elements of local operators and the positive moments of $\phi_B(\tilde{k}_+)$ can indeed be written in terms of local operators. However they diverge as powers of $1/a$ and we still need to develop techniques to subtract these divergences with sufficient precision.

This discussion underlines the fact that we need new theoretical ideas for lattice simulations to start contributing to the evaluation of the non-perturbative QCD effects in $B\to M_1 M_2$ decays and hence to enable fundamental information to be obtained from the wealth of experimental measurements of the rates and CP-asymmetries.

\section{Conclusions}\label{sec:concs}
At this conference we have seen many beautiful contributions to particle physics phenomenology, both in improved precision and in the extensions of computations beyond the standard quantities. We readily forget that it was only a few years ago that results presented at the annual lattice symposium were largely in the quenched approximation, with an error which was not possible to quantify reliably. We then moved on to a brief period with dynamical quarks with masses of $O(500)$\,MeV or so until today we arrive at simulations with almost physical pions. This improvement has to be continued vigorously if precision flavour physics, which has been the focus of this talk, is to play a complementary role to large $p_\perp$ discovery experiments at the LHC in unravelling the next level of fundamental physics. As was noted in sec.\,\ref{sec:nonleptB} however, there are quantities for which a large amount of experimental data is available, and yet for which we do not yet know how to begin formulating the calculations to make them accessible to lattice studies.

At the previous lattice conference which Guido Martinelli helped to organise which was held in 1989 in Capri, Ken Wilson made the seemingly pessimistic prediction that it will take about 30 years to have precision Lattice QCD. We only have 9 years left to fulfill the prediction, but we are now well on our way.

\section*{Acknowledgements} I thank my colleagues from the RBC and UKQCD collaborations with whom many of the ideas and results presented in this talk were developed and obtained and my colleagues from the Flavianet Lattice Averaging Group for many constructively critical discussions placing lattice results in a wider context. I also warmly thank Andrzej Buras for his longstanding encouragement and patient explanations and Guido Martinelli for discussions about the evaluation of long-distance effects. I acknowledge partial support from the STFC Grant ST/G000557/1 and from the EU contract MRTN-CT-2006-035482 (Flavianet).

\end{document}